\documentclass[aps,prd,showpacs,twocolumn,groupedaddress]{revtex4}
\usepackage{graphicx}

\begin{document}

\title{Relevant gluonic energy scale of spontaneous chiral symmetry breaking\\from lattice QCD}

\author{Arata~Yamamoto}
\affiliation{Department of Physics, Faculty of Science, Kyoto University, Kitashirakawa, Sakyo, Kyoto 606-8502, Japan}

\author{Hideo~Suganuma}
\affiliation{Department of Physics, Faculty of Science, Kyoto University, Kitashirakawa, Sakyo, Kyoto 606-8502, Japan}

\date{\today}

\begin{abstract}
We analyze which momentum component of the gluon field induces spontaneous chiral symmetry breaking in lattice QCD.
After removing the high-momentum or low-momentum component of the gluon field, we calculate the chiral condensate and observe the roles of these momentum components.
The chiral condensate is found to be drastically reduced by removing the zero-momentum gluon.
The reduction is about 40\% of the total in our calculation condition.
The nonzero-momentum infrared gluon also has a sizable contribution to the chiral condensate.
From the Banks-Casher relation, this result reflects the nontrivial relation between the infrared gluon and the zero-mode quark.
\end{abstract}


\maketitle

\section{Introduction}
Spontaneous symmetry breaking is one of the most significant and universal mechanisms in physics \cite{Na61,Ko73,We67,Sa68}.
Although the QCD Lagrangian possesses chiral symmetry in the chiral limit, SU($N_f)_L \times$ SU($N_f)_R$ symmetry is spontaneously broken into its subgroup SU($N_f)_V$.
Spontaneous chiral symmetry breaking is remarkably important in hadron physics \cite{Ha94,Hi91,Mi93}.
Also, it is one of the dominant origins of mass in our world.

Chiral symmetry itself is symmetry of quarks, not gluons.
However, spontaneous breaking is dynamically induced by the nonperturbative interaction of gluons.
The gluon dynamics is inseparably linked with chiral symmetry breaking.
Our goal is to determine {\it what momentum component of the gluon field induces spontaneous chiral symmetry breaking}.
The relation between the eigenmode of quarks and chiral symmetry breaking is known as the Banks-Casher relation \cite{Ba80}.
On the other hand, the relation between the momentum component of gluons and chiral symmetry breaking is nontrivial.
It is easy to expect the importance of the low-momentum gluon, but difficult to predict the detailed relation due to the nonperturbative dynamics of the low-momentum gluon.
We would like to clarify such a relation nonperturbatively by lattice QCD.
In other words, we quantitatively investigate {\it the relevant gluonic energy scale} of spontaneous chiral symmetry breaking from lattice QCD.

To analyze the relevant gluonic energy scale, we consider momentum space of the gluon field.
The gluon field is described by the link variable in lattice QCD.
By manipulating the link variable in momentum space, we directly analyze how the momentum component of the gluon field affects chiral symmetry breaking.
The obtained energy scale would also be interesting from the viewpoint of a connection to other QCD phenomena, such as confinement \cite{Ma84,Su95,Mi95,Mo04}.
In lattice QCD, the connection between confinement and chiral symmetry breaking is investigated in the context of phase transition at finite temperature \cite{Ko83,Po84,Fu86,Si08}.
Our analysis is a different approach to reveal this connection.

We calculate the chiral condensate $\langle \bar{q}q \rangle$ in lattice QCD.
The chiral condensate is an order parameter of chiral symmetry breaking in the chiral limit.
It is nonzero in the symmetry-broken phase and zero in the symmetry-restored phase.
We denote the flavor-averaged chiral condensate in the lattice unit as
\begin{eqnarray}
\Sigma \equiv - \frac{1}{N_f} a^3 \langle \bar{q}q \rangle = \frac{1}{N_f} a^3 {\rm tr} S_q,
\end{eqnarray}
where $a$ is the lattice spacing and $S_q$ is the quark propagator.
When the quark mass is finite, the chiral condensate includes the effect of explicit breaking by the quark mass as well as spontaneous breaking.
To extract the chiral limit in lattice QCD, one calculates with several quark masses and extrapolates to the chiral limit.

In this paper, we calculate the chiral condensate in SU(3)$_c$ quenched and full lattice QCD, and analyze the relevant gluonic energy scale of spontaneous symmetry breaking.
This paper is organized as follows.
In Sec.~II, we explain how to analyze the relevant gluonic energy scale in lattice QCD.
In Sec.~III, we show the simulation setup of the lattice QCD calculation.
In Sec.~IV, we present the numerical result of the chiral condensate and analyze how the chiral condensate is affected by removing the high-momentum or low-momentum gluon.
Finally, Sec.~V is devoted to a conclusion.

\section{Formalism}
The lattice framework to determine the relevant gluonic energy scale was proposed in Ref.~\cite{Ya08}.
In this framework, after artificially removing some momentum component of link variables, one calculates a physical quantity and observes the role of the removed momentum component.
In doing so, one can determine whether the momentum component is relevant or not for the quantity.
To be self-contained, we briefly introduce the procedure in the following.

Step 1. The SU(3)$_c$ link variable $U_{\mu}(x)$ is generated by Monte Carlo simulation.
As explained below, the link variable must be fixed with a certain gauge.
In this paper, we use the Landau gauge for the numerical calculation.
In the Landau gauge, the gauge fluctuation is minimized and the connection between the link variable and the gauge field is straightforward.

Step 2. The momentum-space link variable ${\tilde U}_{\mu}(p)$ is obtained by the Fourier transformation, as
\begin{eqnarray}
{\tilde U}_{\mu}(p)=\frac{1}{N_{\rm site}}\sum_{x} U_{\mu}(x)\exp(i \sum_{\nu} p_\nu x_\nu ),
\end{eqnarray}
where $N_{\rm site}$ is the total number of lattice sites.

\begin{figure}[t]
\begin{center}
\includegraphics[scale=0.6]{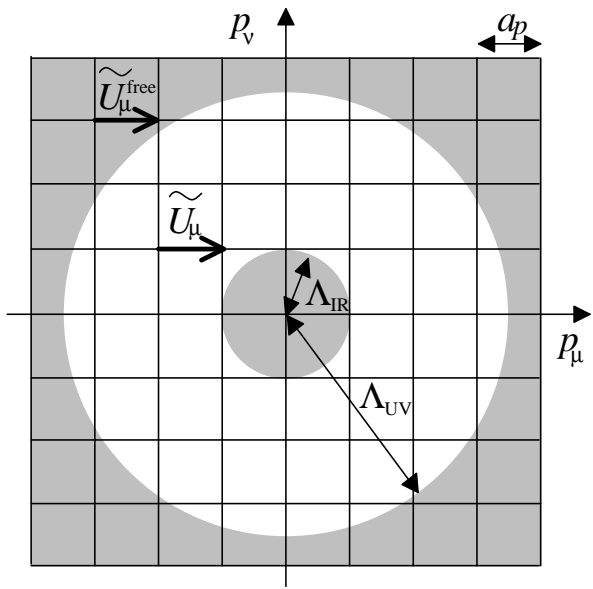}
\caption{\label{fig1}
The schematic figure of momentum space.
The shaded regions are the cut regions by the ultraviolet cutoff $\Lambda_{\rm UV}$ and the infrared cutoff $\Lambda_{\rm IR}$.
The momentum-space lattice spacing is $a_p= 2\pi/La$.
}
\end{center}
\end{figure}

Step 3. Some component of ${\tilde U}_{\mu}(p)$  is removed by introducing a momentum cutoff.
In the cut region, the momentum-space link variable is replaced by the free-field link variable
\begin{equation}
{\tilde U}^{\rm free}_{\mu}(p)=\frac{1}{N_{\rm site}}\sum_x 1 \exp(i {\textstyle \sum_\nu} p_\nu x_\nu)=\delta_{p0}.
\end{equation}
For example, in the case of the ultraviolet cutoff $\Lambda_{\rm UV}$, the momentum-space link variable is replaced as 
\begin{equation}
{\tilde U}_{\mu}^{\Lambda}(p)= \Bigg\{
\begin{array}{cc}
{\tilde U}_{\mu}(p) & (\sqrt{p^2} \le \Lambda_{\rm UV})\\
0 & (\sqrt{p^2} > \Lambda_{\rm UV}).
\end{array}
\end{equation}
In the case of the infrared cutoff $\Lambda_{\rm IR}$, it is replaced as
\begin{equation}
{\tilde U}_{\mu}^{\Lambda}(p)= \Bigg\{
\begin{array}{cc}
\delta_{p0} & (\sqrt{p^2} < \Lambda_{\rm IR})\\
{\tilde U}_{\mu}(p) & (\sqrt{p^2} \ge \Lambda_{\rm IR}).
\end{array}
\end{equation}
The schematic figure is shown in Fig.~\ref{fig1}.

Step 4. The coordinate-space link variable with the momentum cutoff is obtained by the inverse Fourier transformation as
\begin{eqnarray}
U'_{\mu}(x)=\sum_{p} {\tilde U}_{\mu}^{\Lambda}(p)\exp (-i \sum_{\nu} p_\nu x_\nu ).
\end{eqnarray}
Since $U'_{\mu}(x)$ is not an SU(3)$_c$ matrix in general, $U'_{\mu}(x)$ must be projected onto an SU(3)$_c$ element $U^{\Lambda}_{\mu}(x)$.
The projection is realized by maximizing the quantity
\begin{eqnarray}
{\rm ReTr}[\{ U^{\Lambda}_{\mu}(x) \}^{\dagger}U'_{\mu}(x)].
\end{eqnarray}

Step 5. The expectation value of an operator $O$ is computed by using this link variable $U^{\Lambda}_{\mu}(x)$ instead of $U_{\mu}(x)$, i.e., $\langle O[U^\Lambda]\rangle$ instead of $\langle O[U]\rangle$.

Repeating these five steps with various values of the momentum cutoff, we observe the dependence on the momentum cutoff.
Then, we can determine what momentum component of the gluon field is relevant for the physical quantity.
The framework is applicable to both quenched and full QCD in the same way.

Indeed, this framework is powerful in determining the relevant gluonic energy scale of confinement in quenched QCD \cite{Ya08,Ya09}.
By applying this framework to the calculation of the Wilson loop, it was found that the string tension is generated by the infrared gluon below about 1.5 GeV.
By picking up this relevant momentum component, the quark-antiquark potential is clearly decomposed into the confinement potential and the perturbative potential.
Hence, the relevant gluonic energy scale of confinement was determined to be $\sqrt{p^2}\le 1.5$ GeV.

We comment on two points of the framework.
The first is the gauge fixing in Step 1. 
In general, since the gauge transformation is nonlocal in momentum space, the momentum region of the gauge field is a gauge-dependent concept.
Then, our result would depend on the gauge choice.
We show the Landau-gauge results in this paper.
Note, however, that one can analyze other gauges and the gauge dependence since the framework itself does not depend on the gauge choice \cite{Ya08,Ya09}.

The second is the projection in Step 4.
Although such a projection is often used in SU(3)$_c$ lattice QCD as a workable method, the projection could in principle contaminate the original condition on the momentum cutoff.
To evaluate how the projection changes link variables, we calculate $U^{\Lambda\Lambda}_{\mu}(x)$ by adopting Steps 2-4 once again to $U^{\Lambda}_{\mu}(x)$, and check the overlap between them, $\frac{1}{3}{\rm ReTr}[\{ U^{\Lambda}_{\mu}(x) \}^{\dagger}U^{\Lambda\Lambda}_{\mu}(x) ]$. 
The overlap is found to be almost unity.
For example, the deviation from unity is about 0.1\% at $\Lambda_{\rm IR}=1.5$ GeV.
Then, we can expect that the projection does not significantly change link variables.
In fact, we have already reached a steady state configuration with the single procedure.

\section{Simulation setup}

\begin{table}[b]
\caption{\label{tab1}
The parameters of full and quenched lattice QCD configurations.
The dynamical quark mass $m_{\rm sea}$, the configuration number $N_{\rm conf}$, the lattice spacing $a$, and the momentum-space lattice spacing $a_p$ are listed.
}
\begin{tabular}{ccccccc}
\hline\hline
& $\beta$ & Volume & $m_{\rm sea}a$ & $N_{\rm conf}$ & $a$ [fm] & $a_p$ [GeV] \\
\hline
Full &  5.7 & $16^3\times 32$ & 0.01 & 24 - 49 & 0.098 & 0.79 \\
Quenched &  6.0 & $32^4$ & - & 10 & 0.100 & 0.39 \\
\hline\hline
\end{tabular}
\end{table}

The lattice QCD simulations are performed in SU(3)$_c$ quenched and full QCD.
The parameters of gauge configurations are summarized in Table \ref{tab1}.
For the full QCD calculation, we use the dynamical configuration which includes the two-flavor staggered quark in NERSC archive \cite{Br91}.
The momentum-space lattice spacing $a_p$ is given by $a_p\equiv 2\pi/La$, where $L$ is the number of lattice sites in the spatial direction.

To compute the chiral condensate, we adopt the staggered fermion action, which preserves the U(1) subgroup of the full chiral symmetry in the chiral limit.
In full QCD, we use a single mass for the valence and sea quarks, $ma=m_{\rm sea}a=0.010$.
The corresponding pion mass is about 500 MeV and the flavor-averaged chiral condensate is about (540 MeV)$^3$.
In quenched QCD, we use the quark masses $ma=0.010$, 0.015, and 0.025 to extrapolate the chiral limit.

\section{lattice QCD result}
\subsection{Chiral condensate with the UV cutoff}

\begin{figure}[t]
\begin{center}
\includegraphics[scale=1.2]{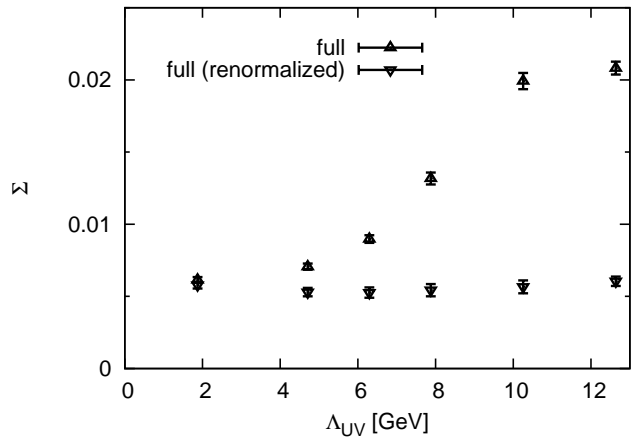}
\caption{\label{fig2}
The chiral condensate $\Sigma \equiv -a^3\langle \bar{q}q \rangle/N_f$ with the UV cutoff $\Lambda_{\rm UV}$
The quark mass is $ma=0.01$.
The ``renormalized" chiral condensate is multiplied by the renormalization factor $Z_S$.
}
\end{center}
\end{figure}

First, we show the chiral condensate with the ultraviolet (UV) cutoff $\Lambda_{\rm UV}$ in Fig.~\ref{fig2}.
Since there is no significant difference between the quenched and full QCD results, we plot only the full QCD result.
The right-side point at $\Lambda_{\rm UV}\simeq 12.5$ GeV is the result of original lattice QCD without the momentum cutoff.

Although spontaneous chiral symmetry breaking is expected to be caused by nonperturbative gluons, the chiral condensate is drastically changed by the UV cutoff.
However, as shown below, this is mainly because the chiral condensate is a renormalization-group variant and UV-diverging quantity.
It is dressed by perturbative gluons and its value strongly depends on the UV regularization.
In standard lattice QCD, the perturbative contribution is several orders of magnitude larger than the nonperturbative core of the condensate \cite{YaS}.

To estimate the effect of renormalization, we calculate a renormalization factor, so-called a Z-factor, nonperturbatively \cite{Ma95,Ao99}.
The renormalization factor $Z_O(k)$ is determined from the amputated Green function of the quark bilinear operator $O$.
The renormalization condition is imposed as
\begin{eqnarray}
Z_O (k) Z^{-1}_q (k) \Gamma_O (k)=1,
\end{eqnarray}
where
\begin{eqnarray}
\Gamma_O (k) &\equiv& \frac{1}{16N_c} {\rm tr} [ S_q^{-1}(k) G_O (k) S_q^{-1}(k) P^\dagger_O]\\
G_O (k) &\equiv& \langle q(k) O \bar{q}(k) \rangle \\
S_q(k) &\equiv& \langle q(k) \bar{q}(k) \rangle.
\end{eqnarray}
$P_O$ is the appropriate projection operator.
The wave-function renormalization factor $Z_q^{1/2}(k)$ of the quark field is obtained from the conserved vector current, i.e., $Z_V(k)=1$.
Note that $k$ is the momentum of the quark field, not the momentum of the gluon field.

We calculate the renormalization factor $Z_S(k)$ of the scalar operator, and plot the renormalized chiral condensate $Z_S(5\ {\rm GeV}) \times \Sigma$ in Fig.~\ref{fig2}.
The renormalized chiral condensate is almost independent of the UV cutoff.
As the UV gluon is removed by the UV cutoff, the bare chiral condensate approaches the renormalized one.
This means that the drastic change by the UV cutoff is well explained in terms of renormalization.

\subsection{Chiral condensate with the IR cutoff}

\begin{figure}[t]
\begin{center}
\includegraphics[scale=1.2]{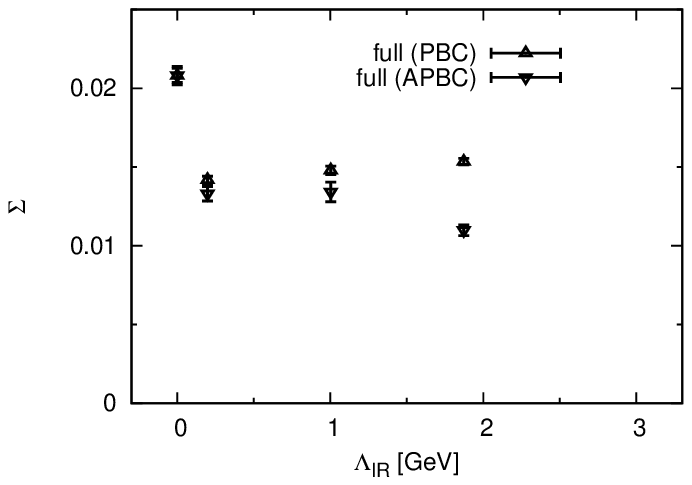}
\caption{\label{fig3}
The full QCD result of the chiral condensate $\Sigma \equiv -a^3\langle \bar{q}q \rangle/N_f$ with the IR cutoff $\Lambda_{\rm IR}$.
The lattice volume is $16^3 \times 32$, and the quark mass is $ma=0.01$.
PBC and APBC mean periodic and antiperiodic boundary conditions, respectively.
}
\end{center}

\begin{center}
\includegraphics[scale=1.2]{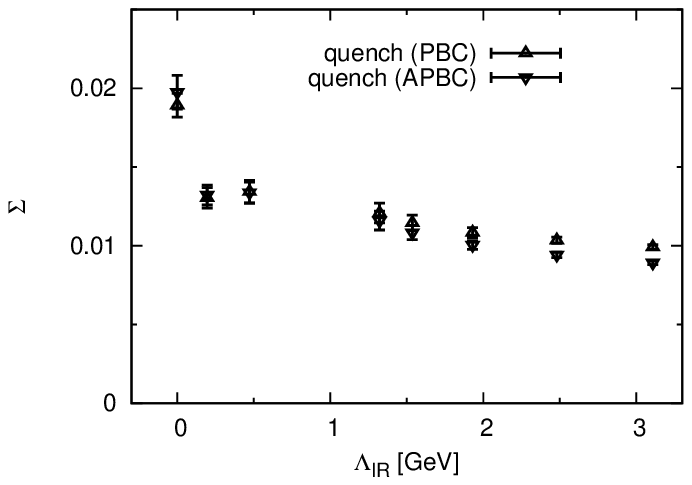}
\caption{\label{fig4}
The quenched QCD result of the chiral condensate with the IR cutoff.
The lattice volume is $32^4$, and the quark mass is $ma=0.01$.
The notation is the same as Fig.~\ref{fig3}.
}
\end{center}
\end{figure}

Second, we analyze the chiral condensate with the infrared (IR) cutoff $\Lambda_{\rm IR}$.
We show the full QCD result in Fig.~\ref{fig3} and the quenched QCD result in Fig.~\ref{fig4}.
The quark mass is $ma=0.01$ in both calculations.
In the case of the IR cutoff, the chiral condensate does not show the drastic change corresponding to renormalization.
Then, we expect the physical contribution to spontaneous chiral symmetry breaking instead of an artifact of renormalization.

When the IR gluon is removed, the effective quark mass would be reduced especially  at a large distance.
Thus, we must pay attention to the finite-volume effect in $\Lambda_{\rm IR}>0$, even though our lattice volume is large enough at $\Lambda_{\rm IR}=0$.
We estimate the finite-volume effect by changing boundary conditions of the quark propagator \cite{Do04}.
In Fig.~\ref{fig3} and Fig.~\ref{fig4}, PBC and APBC mean periodic and antiperiodic boundary conditions, respectively.
Since the result is independent of the boundary conditions if the lattice volume is large enough, the difference between these data should be understood as the finite-volume effect.
As seen from Fig.~\ref{fig3}, the $16^3 \times 32$ lattice of full QCD suffers from the finite-volume effect in $\Lambda_{\rm IR}>1.0$ GeV.
From Fig.~\ref{fig4}, the finite-volume effect is fairly small for the $32^4$ lattice of quenched QCD,
although it gradually grows in $\Lambda_{\rm IR}>1.5$ GeV.

Both in Fig.~\ref{fig3} and Fig.~\ref{fig4}, the chiral condensate suddenly gets small around $\Lambda_{\rm IR}=0$.
This jump around $\Lambda_{\rm IR}=0$ is caused by cutting only the zero-momentum link variable ${\tilde U}_{\mu}(0)$.
Despite the change at a single point $p^2=0$, the chiral condensate is about 40\% reduced.
Such a large change is not observed in removing other low-momentum components.
Therefore, the zero-momentum gluon is special and it possesses a major contribution to the chiral condensate.
Note that ``zero momentum" on momentum-space lattice corresponds to the deep-infrared region which is roughly $\sqrt{p^2}<a_p$ in the continuum.

In large $\Lambda_{\rm IR}$, since the lattice volume of full QCD is not large enough, we analyze the quenched QCD result in Fig.~\ref{fig4}.
When the ``nonzero-momentum" gluon of $\sqrt{p^2}\ge a_p$ is removed by the IR cutoff, the chiral condensate gradually decreases.
Thus, not only the ``zero-momentum" gluon but also the ``nonzero-momentum" gluon contributes to the chiral condensate.
The chiral condensate continues to decrease even in $\Lambda_{\rm IR}>1.5$ GeV.
Although it is difficult to perform an accurate analysis in large $\Lambda_{\rm IR}$ due to the finite-volume effect, we can see that the chiral condensate is also affected by the gluon in the intermediate-momentum region.

\subsection{Chiral extrapolation}

\begin{figure}[t]
\begin{center}
\includegraphics[scale=1.2]{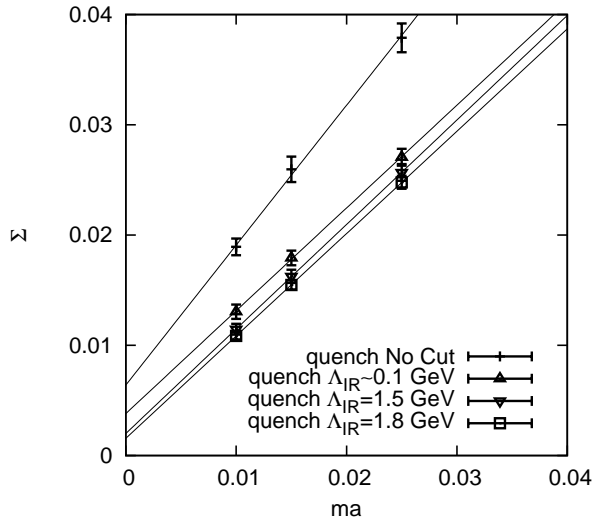}
\caption{\label{fig5}
The chiral extrapolation of the chiral condensate $\Sigma \equiv -a^3\langle \bar{q}q \rangle/N_f$.
$\Lambda_{\rm IR}\sim 0.1$ GeV corresponds to the cutoff for the zero-momentum link variable.
}
\end{center}
\end{figure}

Next, we consider the chiral extrapolation of the chiral condensate.
When the bare quark mass $m$ is small, the chiral condensate is expanded as a function of $m$, as
\begin{eqnarray}
\Sigma (m) = \Sigma (0) + ma \Sigma'(0) + \cdots,
\end{eqnarray}
where $\Sigma'(m) \equiv \partial \Sigma (m)/ \partial ma$.
$\Sigma (0)$ represents spontaneous chiral symmetry breaking in the chiral limit.
We fit the quenched QCD result by the linear extrapolation function $\Sigma (0) + ma \Sigma'(0)$.
The fitting result is shown in Fig.~\ref{fig5} and Table \ref{tab2}.
Note that the data of ``$\Lambda_{\rm IR}\sim 0.1$ GeV" corresponds to the smallest IR cutoff, which cuts only the zero-momentum link variable, and so the value ``0.1 GeV" itself is not so meaningful.

As stated above, when the zero-momentum gluon field is removed, the chiral condensate is largely changed.
$\Sigma (0)$ is about 40\% reduced and $\Sigma'(0)$ is about 30\% reduced.
As for the nonzero-momentum gluon, the extrapolating line moves down parallel by the infrared cutoff.
$\Sigma (0)$ is gradually reduced and $\Sigma'(0)$ is almost unchanged.
This indicates that the nonzero-momentum gluon has small but finite contribution to spontaneous chiral symmetry breaking.

\begin{table}[b]
\caption{\label{tab2}
The fitting result of the chiral extrapolation in Fig.~\ref{fig5}.
The extrapolation function is $\Sigma (0) + ma \Sigma'(0)$.
}
\begin{tabular}{ccc}
\hline\hline
$\Lambda_{\rm IR}$ & $\Sigma (0)$ & $\Sigma'(0)$\\
\hline
0       & 0.00639(81) & 1.269(53)\\
$\sim 0.1$ GeV & 0.00380(27) & 0.933(16)\\
1.5 GeV & 0.00200(7)  & 0.948(4) \\
1.8 GeV & 0.00155(2)  & 0.929(1) \\
\hline\hline
\end{tabular}
\end{table}

In Fig.~\ref{fig5}, our result suggests another interesting possibility.
At least within the present numerical accuracy, the chiral condensate in the chiral limit remains finite at $\Lambda_{\rm IR}=1.5$ GeV, which is the relevant gluonic energy scale of confinement.
If this is true, this means that the gluonic energy scale of spontaneous chiral symmetry breaking is larger than that of color confinement at zero temperature.
Unfortunately, however, we cannot make a decisive statement due to systematic error of the chiral extrapolation.
For more conclusive answer, we need the full QCD calculation very close to the chiral limit, while the finite-volume effect is severely crucial in large $\Lambda_{\rm IR}$ and small $m$.

\section{Conclusion}
We have analyzed which momentum component of the gluon field induces spontaneous chiral symmetry breaking.
The most dominant contribution is given by the ``zero-momentum" gluon, which roughly corresponds to the deep-infrared region of $\sqrt{p^2}<a_p$ in the continuum.
Not only zero-momentum but also the ``nonzero-momentum" gluon of $\sqrt{p^2}>a_p$ possesses a sizable contribution.
While we cannot precisely determine the upper limit of the relevant momentum component of the gluon field due to the finite-volume effect, its relevant momentum component seems to be broadly distributed to the intermediate-momentum region.

The zero-momentum gauge field corresponds to a spatially-uniform gauge background.
In general, the non-Abelian gauge field could have a nontrivial effect even in spatially-uniform case, unlike the Abelian gauge field. 
Our result actually suggests that the zero-momentum gauge field contributes to the chiral condensate. 
Note, however, that it is nontrivial whether spontaneous chiral symmetry breaking occurs only by the spatially-uniform gauge background. 

The Banks-Casher relation states that the chiral condensate is related to the spectral density $\rho (\lambda)$ of the Dirac operator as
\begin{eqnarray}
\langle \bar{q}q \rangle = - \pi \rho (0)
\end{eqnarray}
in the chiral limit \cite{Ba80}.
The spectral density of the Dirac operator is given in infinite volume as
\begin{eqnarray}
\rho (\lambda) = \lim _{V\to \infty} \frac{1}{V} \sum_k \delta (\lambda -\lambda_k),
\end{eqnarray}
and the eigenvalue of the Dirac operator is $i\lambda_k$.
The zero mode of quarks is directly related to spontaneous chiral symmetry breaking from this relation.
In contrast, the gluon field is nontrivially related to spontaneous chiral symmetry breaking.
Our result presents the connection between the momentum component of gluons and the zero mode of quarks.

Although the relation between the energy scales of confinement and chiral symmetry breaking is interesting, our result is not conclusive but suggestive in the present accuracy.
To approach the realistic situation of QCD, we would need the reliable chiral extrapolation including the dynamical quark effect.

\section*{Acknowledgements}
A.~Y.~and H.~S.~are supported by a Grant-in-Aid for Scientific Research [(C) No.~20$\cdot$363 and (C) No.~19540287] in Japan.
The authors are grateful to Dr.~T.~Doi for the quark solver code.
The lattice QCD calculations are done on NEC SX-8R at Osaka University.
The full QCD gauge configuration is provided from NERSC archive.
This work is supported by the Global COE Program, ``The Next Generation of Physics, Spun from Universality and Emergence," at Kyoto University.


\begin{thebibliography}{99}
\bibitem{Na61} Y.~Nambu and G.~Jona-Lasinio, Phys. Rev. {\bf 122}, 345 (1961); Phys. Rev. {\bf 124}, 246 (1961).
\bibitem{We67} S.~Weinberg, Phys. Rev. Lett. {\bf 19}, 1264 (1967).
\bibitem{Sa68} A.~Salam, {\it Elementary Particle Theory} (Almqvist and Wiksell, Stockholm, 1968).
\bibitem{Ko73} M.~Kobayashi and T.~Maskawa, Prog. Theor. Phys. {\bf 49}, 652 (1973).
\bibitem{Hi91} K.~Higashijima, Prog. Theor. Phys. Suppl. {\bf 104}, 1 (1991). 
\bibitem{Mi93} V.~A.~Miransky, {\it Dynamical Symmetry Breaking in Quantum Field Theories} (World Scientific, Singapore, 1993). 
\bibitem{Ha94} T.~Hatsuda and T.~Kunihiro, Phys.~Rept. {\bf 247}, 221 (1994).
\bibitem{Ba80} T.~Banks and A.~Casher, Nucl. Phys. {\bf B169}, 103 (1980).
\bibitem{Ma84} A.~Manohar and H.~Georgi, Nucl. Phys. {\bf B234}, 189 (1984).
\bibitem{Su95} H.~Suganuma, S.~Sasaki, and H.~Toki, Nucl. Phys. {\bf B435}, 207 (1995).
\bibitem{Mi95} O.~Miyamura, Phys. Lett. {\bf B353}, 91 (1995).
\bibitem{Mo04} A.~Mocsy, F.~Sannino, and K.~Tuominen, Phys. Rev. Lett. {\bf 92}, 182302 (2004).
\bibitem{Ko83} J.~B.~Kogut, M.~Stone, H.~W.~Wyld, W.~R.~Gibbs, J.~Shigemitsu, S.~H.~Shenker, and D.~K.~Sinclair, Phys. Rev. Lett. {\bf 50}, 393 (1983).
\bibitem{Po84} J.~Polonyi, H.~W.~Wyld, J.~B.~Kogut, J.~Shigemitsu, and D.~K.~Sinclair, Phys. Rev. Lett. {\bf 53}, 644 (1984).
\bibitem{Fu86} M.~Fukugita and A.~Ukawa, Phys. Rev. Lett. {\bf 57}, 503 (1986).
\bibitem{Si08} D.~K.~Sinclair, Phys. Rev. D {\bf 78}, 054512 (2008).
\bibitem{Ya08} A.~Yamamoto and H.~Suganuma, Phys. Rev. Lett. {\bf 101}, 241601 (2008).
\bibitem{Ya09} A.~Yamamoto and H.~Suganuma, Phys. Rev. D {\bf 79}, 054504 (2009).
\bibitem{Br91} F.~R.~Brown {\it et al.}, Phys. Rev. Lett. {\bf 67}, 1062 (1991).
\bibitem{YaS} A.~Yamamoto, arXiv:0906.2618 (2009).
\bibitem{Ma95} G.~Martinelli, C.~Pittori, C.~T.~Sachrajda, M.~Testa, and A.~Vladikas, Nucl. Phys. {\bf B445}, 81 (1995).
\bibitem{Ao99} S.~Aoki {\it et al.} (JLQCD Collaboration), Phys. Rev. Lett. {\bf 82}, 4392 (1999).
\bibitem{Do04} T.~Doi, N.~Ishii, M.~Oka, and H.~Suganuma, Phys. Rev. D {\bf 70}, 034510 (2004).
\end{thebibliography}
\end{document}